\documentclass[12pt,a4paper]{article}
\usepackage{amsmath, amssymb}
\usepackage{mathtools}
\usepackage{upgreek}
\usepackage{graphicx}
\usepackage[T1]{fontenc}
\usepackage[utf8]{inputenc}
\usepackage{authblk}
\newcommand{\be}{\begin{equation}}
\newcommand{\ee}{\end{equation}}
\newcommand{\bear}{\begin{eqnarray}}
\newcommand{\eear}{\end{eqnarray}}
\title{\textbf{On the Independence of Electromagnetism and Gravitation from Test Bodies Physical Properties}}
\author{Clovis Jacinto de Matos\footnote{\textit{clovis.de\_matos@mailbox.tu-dresden.de}}}

\affil{ Institute of Aerospace Engineering, Technische Universit\"{a}t Dresden, Marschnerstrasse 32, 01307 Dresden, Germany}
\date{\today}

\begin{document}
\maketitle
\begin{abstract}
Does a physical field exist independently of the interaction between the field source and the test body used to measure it at a given point? What does propagate between two physical bodies when they interact? These are the fundamental questions we answer in the present work for electromagnetic and gravitational interactions. We come to the conclusion that gravitational fields as well as spacetime curvature cannot depend on the passive-gravitational to inertial mass ratio of the test bodies, because this would lead to energy conservation and causality violation except when this ratio is universally exactly equal to one. For what concerns the hypothetical dependence of electromagnetic fields on the charge to inertial-mass ratio of test bodies, this would call for the propagation of an operationally ill defined quantity (Coulomb . Meter$^2$ / Second$^2$) that would replace energy (Joule = Kilogram . Meter$^2$ / Second$^2$) transfer in the interaction between two electrically charged bodies. Hence electromagnetic fields cannot depend on the charge to inertial-mass ratio of test bodies. A consequence of these results is also the disqualification of Murat Ozer's unification theory of gravitation and electromagnetism.
\begin{description}
\item[Key words]
Field, test body, interaction, action at a distance, locality
\end{description}
\end{abstract}

\section{Introduction}
Prof. Percy Williams Bridgman questioned radically the operational fundamental physical nature of gravitational and electromagnetic fields \cite{Bridg1} \cite{Bridg2}. In his view these are only mathematical constructs and their physical reality must be found in the forces experienced by test bodies in the vicinity of a gravitational or electromagnetic source. This operational approach of the electromagnetic and gravitational fields, forms the foundation of Prof. Murat Ozer theory of unification of electromagnetism with gravitation \cite{MO1} \cite{MO2} \cite{MO3}.

After exposing in detail Bridgman's theory of electromagnetism and gravitation we demonstrate why his approach is incorrect. Leading to operationally ill defined physical quantities, and violations of energy and causality. This has an immediate interest for Ozer's unification theory, which is based on Bridgman's operational criticism of the concept of physical fields.

\section{The physical nature of fields according to Percy Williams Bridgman}\label{sec:bridge}
What is the fundamental nature of physical fields? For Percy Bridgman \cite{Bridg1} "the only physical evidence we ever have of the existence of a field (at a point) is obtained by going there with an electric charge and observing the action on the charge, which is precisely the operation of the definition (of electric field)", i.e "to determine the electric field at a point, we place an exploring charge at the point, measure the force on it, and then calculate the ratio of force to charge." Hence Bridgman, arrives at the conclusion that "the electric field is conceptualized as something characteristic of the point alone, from which any effect which the test charge may have exerted has disappeared with the vanishing of the charge. But this is an improper conceptualization of this limiting process, because force as well as charge vanishes, and the indispensable role of the (test) charge by no means disappears." "...one observes only the way in which a force which at one moment acts on a test-body acts a moment later on another test body". Applying this conclusion to the case of an electromagnetic wave, Bidgman states "The electromagnetic field itself is an invention, and is never subject to direct observation. What we observe are material bodies, with or without charges, their positions, motions, and the forces to which they are subject" cf. pg 136 of \cite{Bridg2} (see also other indicated pages in this bibliographic reference for additional information on Bridgman's views about the concept of electromagnetic field).

In summary, for Bridgman, physical fields, like gravitational, or electric or magnetic fields, do not exist independently of the forces measured on the test body.The fundamental physical nature of a field would be, according to Bridgman, the acceleration observed on the test body. The interaction between the source of the field and the test body, reduces to the acceleration, supported by the test body, measured at the point where the test body is located with respect to the source. This concept of interaction does not contain any, self standing, independent, physical entity (like a field) propagating from the source of the field to the test body. Hence the measured acceleration of the test body would be the result of a "mysterious team work" between the source of the field and the test body. To state it blandly, the interaction between the field source and the test body would not be based on the propagation of a field, but instead on a Newtonian-kind action-at-a-distance. The effect of the field source on the test body being instantaneous (i.e. propagating at an infinite speed).

\begin{figure}[htb]
\begin{center}
\includegraphics[width=\textwidth]{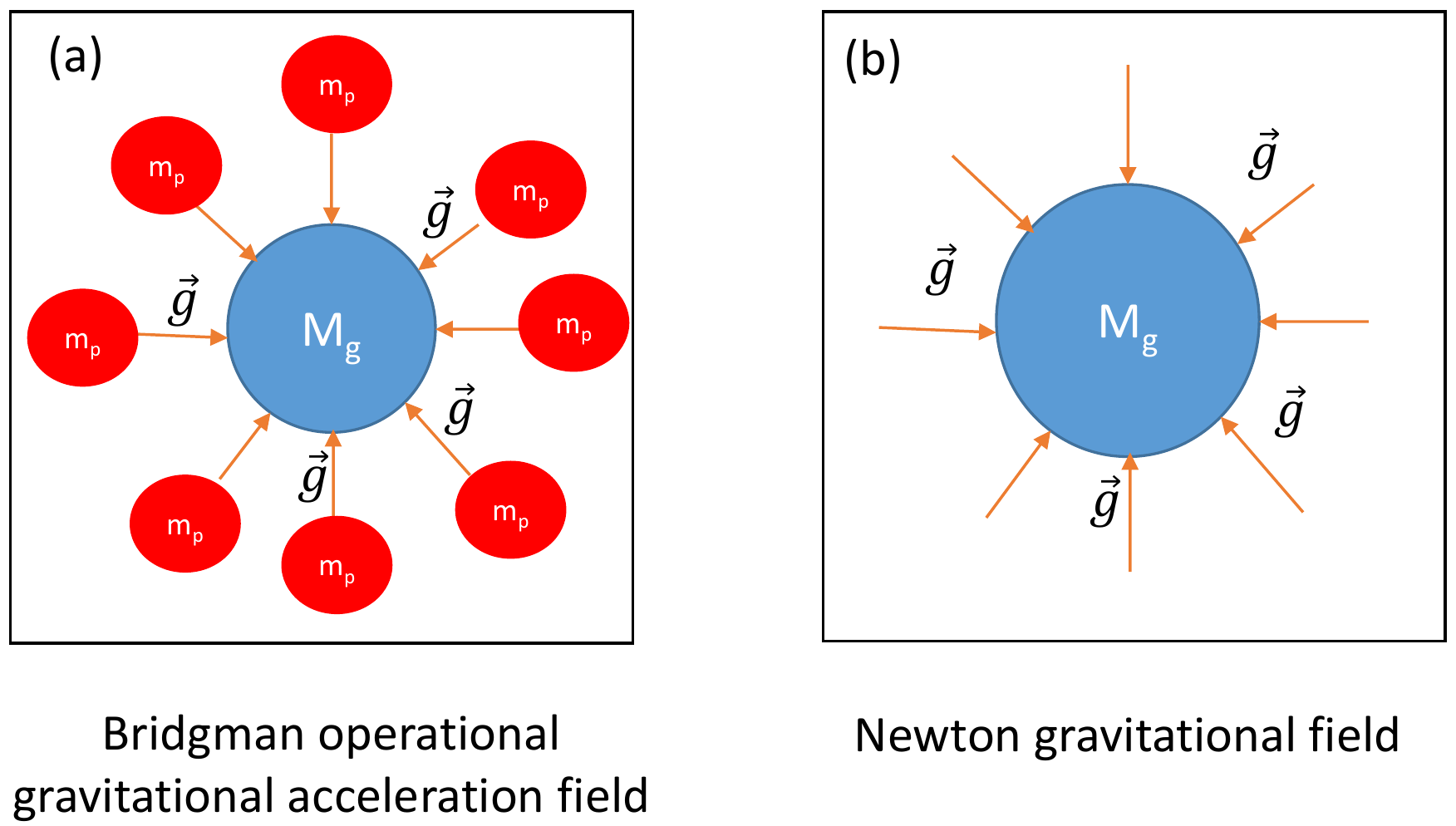}
\end{center}
\caption{According to Bridgman, the gravitational acceleration $\vec g$ is caused by the combined presence of the active-gravitational mass of the source, $M_g$, and the passive-gravitational mass of the test body $m_p$, cf. inset (a). In inset (b) is shown the Newtonian gravitational field, which is based on a pre-existing gravitational acceleration, $\vec g$, independent of the presence of any test body, and originating only from the gravitational source $M_g$.}\label{fig:gfield}
\end{figure}

Let us make Bridgman's concept of physical interaction more precise by applying it first to a gravitational source consisting in a spherical homogeneous active-gravitational mass $M_g$, and a test body with passive-gravitational mass $m_p$ and inertial mass $m_i$, located at a distance $r$ from the center of gravity of the mass $M_g$, cf. Fig. (\ref{fig:gfield}). The test body will experience the acceleration $\vec g$, which can be estimated from Newton's fundamental law of dynamics:
\begin{equation}
m_i \vec g=\vec F_g \, , \label{bri1}
\end{equation}
where $\vec F_g$ is Newton's universal gravitational law:
\begin{equation}
\vec F_g = - G \frac{m_p M_g}{r^2} \hat{r}\, , \label{bri2}
\end{equation}
where $\hat r$ is the unit vector pointing in the outside direction with respect to $M_g$ mass center. Substituting equation (\ref{bri2}) in equation (\ref{bri1}) we obtain the gravitational acceleration of the test body:
\begin{equation}
\vec g= -\Big(G\frac{m_p}{m_i}\Big) \frac{M_g}{r^2} \hat r \, . \label{bri3}
\end{equation}
Using the following gravitational parameter, $G_N$, instead of the Gravitational universal constant $G$:
\begin{equation}
G_N=G\frac{m_p}{m_i} \, , \label{bri4}
\end{equation}
we can write equation (\ref{bri3}) as:
\begin{equation}
\vec g = - G_N \frac{M_g}{r^2} \hat r \, . \label{bri5}
\end{equation}
We note that even if the passive-gravitational to inertial mass ratio of the test body is equal to one, $m_p/m_i=1$, we cannot forget that the physical meaning of $G_N$ is quite different from the physical nature of the Gravitational universal constant $G$. Newton's gravitational constant is designated as being universal because it is set as being independent of any physical properties of the gravitational source or of the test body. However the acceleration operationally measured by the experimental physicist, involves the term $G_N$, which is assumed to be a universal constant by admitting that the ratio $m_p/m_i=1$ for any test body. Even if experimental tests to date reveal that this is indeed the case \cite{Basseler} \cite{Smith} , this "minor" detail should not be forgotten under the carpet, because it also means that the test body could play an active role in creating the acceleration that it experiences in the presence of the gravitational source $M_g$. Following BRidgman, without the presence of a test body, the physicist could not say anything about the possible influence of the gravitational source $M_g$ on its physical environment.

\begin{figure}[htb]
\begin{center}
\includegraphics[width=\textwidth]{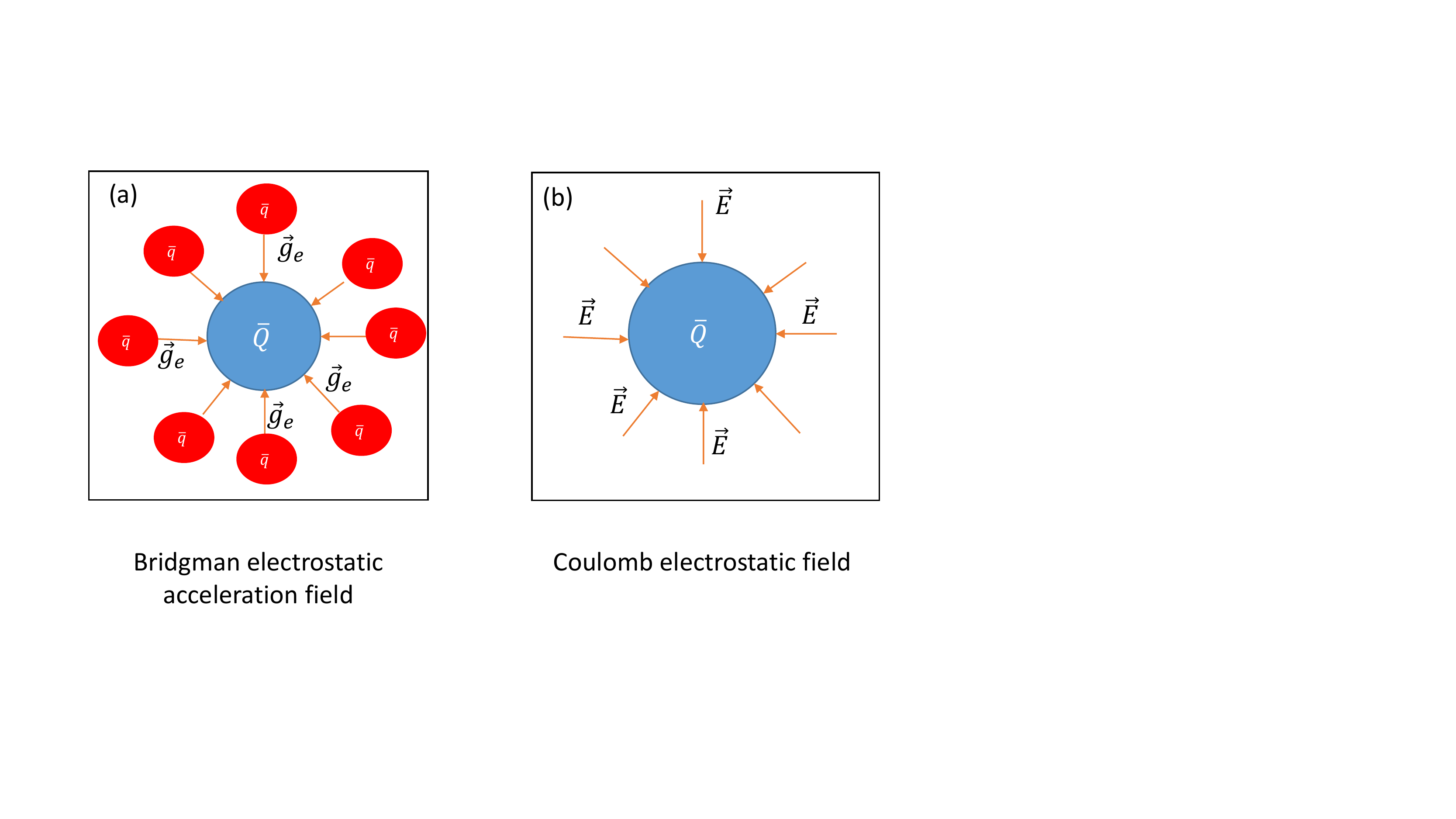}
\end{center}
\caption{Following Bridgman the electric acceleration $\vec g_e$ is caused by the combined presence of the central electric charge $\bar Q=-Q$, and the electric-test mass $\bar q=+q$, cf. inset (a). In inset (b) is represented the Coulomb electrostatic field $\vec E$, that depends only on the electric charge of the source, and is completely independent from the test body electric charge to inertial mass ratio $\bar q/m_i$.}\label{fig:efield}
\end{figure}

The application of Bridgman's operational concept of physical field to the case of the electric field, follows closely what we just did for the gravitational field. We start by using Newton's fundamental law of dynamics to estimate the acceleration, $g_e$, of a test body, with inertial mass $m_i$ and electric charge $\bar q$, in the presence of an electric spherical homogeneous charge $\bar Q$:
\begin{equation}
\vec F_C = m_i \vec g_e\, , \label{bri6}
\end{equation}
where $\vec F_C$ is the Coulomb electrostatic force:
\begin{equation}
\vec F_C=K \frac {\bar q \bar Q}{r^2} \hat r \, , \label{bri7}
\end{equation}
where $K$ is Coulomb constant, $r$ is the distance between the mass centers of charges $\bar Q$ and $\bar q$, and $\hat r$ is a unit vector pointing from $\bar Q$ to $\bar q$. substituting equation (\ref{bri7}) in equation (\ref{bri6}) we obtain:
\begin{equation}
\vec g_e=\Big(K \frac {\bar q}{m}\Big) \frac {\bar Q}{r^2} \hat r \, , \label{bri8}
\end{equation}
Using the following Coulomb parameter instead of Coulomb constant:
\begin{equation}
K_e = K\frac{\bar q}{m_i} \, , \label{bri9}
\end{equation}
we can write equation (\ref{bri8}) in the form:
\begin{equation}
\vec g_e=K_e \frac {\bar Q}{r^2} \hat r \, , \label{bri10}
\end{equation}
Since the electrostatic-acceleration field $\vec g_e$ depends on the charge to inertial-mass ratio of the test body, it was decided to discard this field, in the equations of electrodynamics, and use instead the electrostatic field $\vec E$, which depends only on the electric charge of the electric source:
\begin{equation}
\vec E= K  \frac {\bar Q}{r^2} \hat r \, .\label{bri11}
\end{equation}
However, according to Bridgman, at operational level, the electrostatic field $\vec E$ is never measured directly, it is a quantity always inferred from the measurement of the acceleration $\vec g_e$ at the point where the test charge is present. Although $\vec E$ is a good and useful construct because there is a one to one correspondence between the electric field and the electric charge of the source, let us not discard for the moment the test body acceleration $\vec g_e$ as being the real physical entity replacing the electric field $\vec E$. Substituting equation (\ref{bri11}) in equation (\ref{bri8}) we obtain
\begin{equation}
\vec g_e= \frac {\bar q}{m_i} \vec{E}\, . \label{bri12}
\end{equation}
For Bridgman, an electric charge source and an electric-test mass generate (together) a real acceleration $\vec g_e$, cf. Fig. (\ref{fig:efield}), replacing the electric field, that would not be physically real.
\begin{figure}[htb]
\begin{center}
\includegraphics[width=\textwidth]{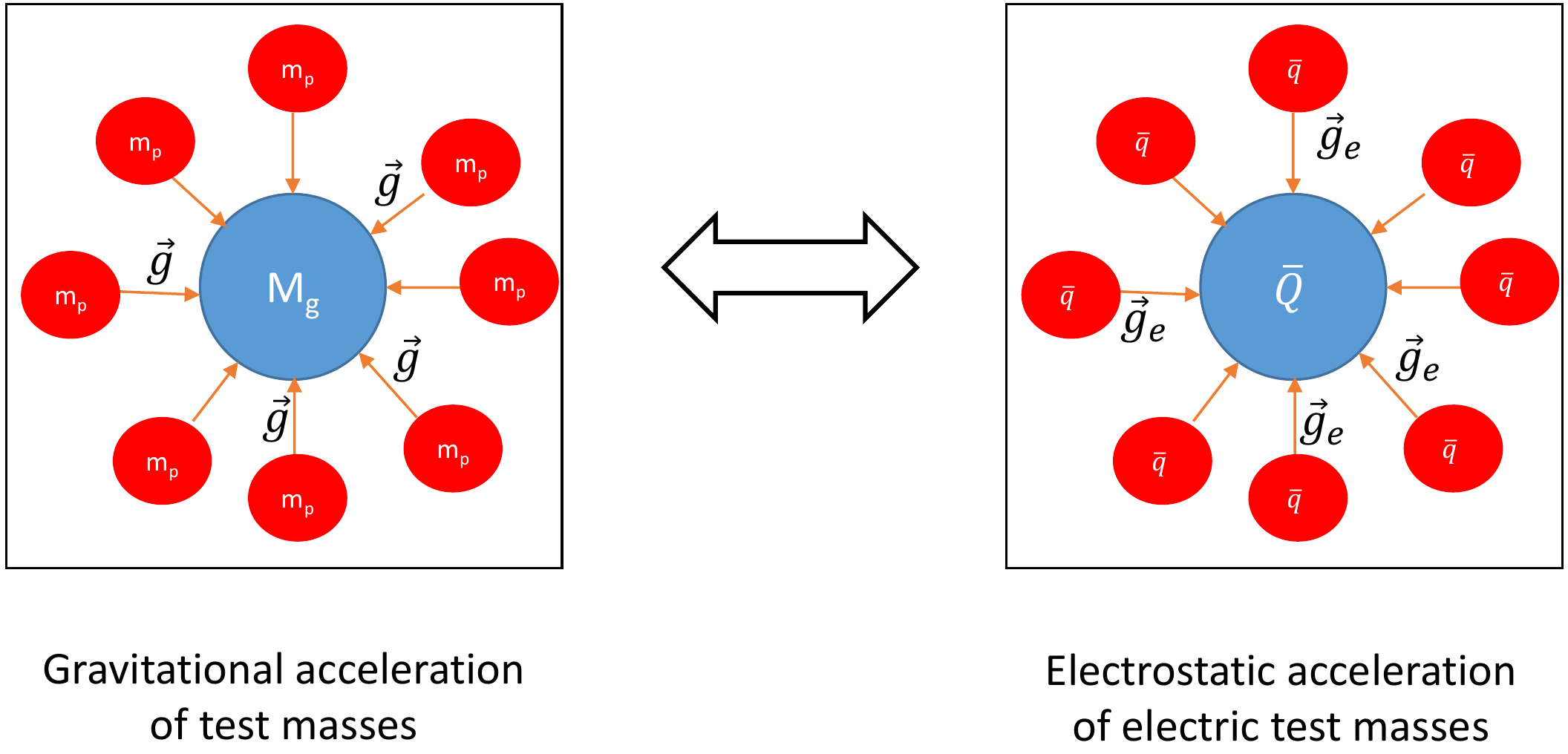}
\end{center}
\caption{According to Bridgman a gravitational acceleration $\vec g$ is physically equivalent to an electrostatic acceleration $\vec g_e$, to the extent that both require the presence of a field source and test bodies to be physically real.}\label{fig:egfield}
\end{figure}

Although $\bar q/m_i$ can have values different from one, and has physical units, contrary to the ratio $m_p/m_i$, which is considered to be always equal to one, and has no units, we still can consider that the gravitational "parameter" $G_N$ is equivalent to the electric parameter,  $K_e$, to the extent that both parameters reflect the hypothetical active role played by the test body in making the gravitational-acceleration and the electric-acceleration, supported by the test body, possible. These two different accelerations would be equivalent to each other, in the sense that they are both dynamically measured accelerations, replacing the mathematical concepts of gravitational and electric fields  cf. Fig. (\ref{fig:egfield}). In the next section, we demonstrate that this approach to gravitation and electromagnetism, leads to serious physical inconsistencies.
\section{Electromagnetic waves cannot depend on the test body's charge to inertial-mass ratio}\label{sec:discussion}
Due to the active role of electric test masses in creating the electromagnetic field, as discussed by Bridgman, an electromagnetic wave is not only the propagation of electric and magnetic fields, $E$ and $B$ respectively, oscillating transversely with respect to the propagation direction of the wave, but it is also the transversal propagation of acceleration and gravitomagnetic fields (inertial-coriolis fields) $g_e=\bar q E/m_i$ and $B_g=\bar q B/ 2m_i$ respectively. Although electric and magnetic fields are independent of the physical properties of the electric test masses, acceleration and gravitomagnetic fields, associated by Bridgman with electromagnetic fields, are function of the charge to mass ratio of the test particle.

\begin{figure}[htb]
\begin{center}
\includegraphics[width=\textwidth]{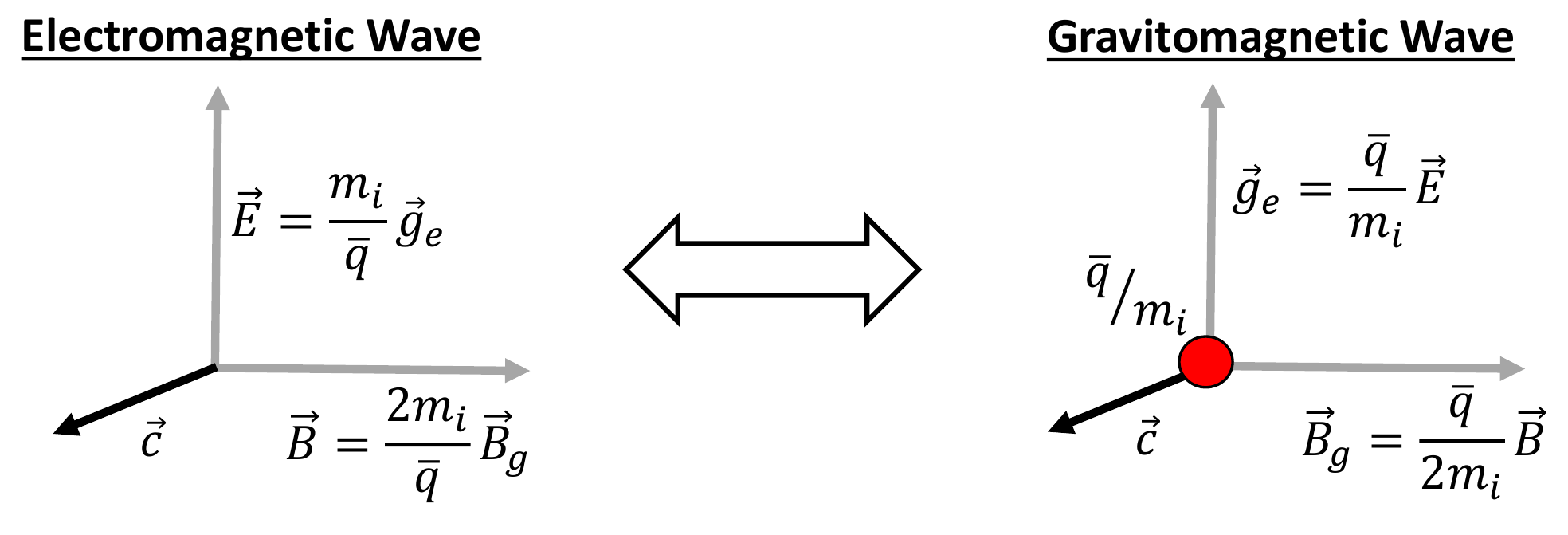}
\end{center}
\caption{Maxwell's electromagnetic waves are represented by Bridgman as (inertial) gravitomagnetic waves. In the picture $\bar q$ and $m_i$ are the electric charge and inertial mass of the test particle respectively, $\vec E$ and $ \vec B$ are the electric and magnetic fields carried out by the electromagnetic wave respectively, $\vec g_e$ and $\vec B_{g}$ are the electric acceleration and the gravitomagnetic (angular rotation) fields supported by the electric-test mass.}\label{fig:gemwave}
\end{figure}

We write Maxwell equations:
\begin{eqnarray}
\nabla \overrightarrow{E} &=&\frac{1}{\epsilon_0}\rho \label{emax}\\
\nabla \overrightarrow{B} &=&0  \nonumber \\
\nabla \wedge \overrightarrow{E} &=&-\frac{\partial \overrightarrow{B}}{
\partial t}  \nonumber \\
\nabla \wedge \overrightarrow{B} &=&\mu_0 \rho
\overrightarrow{v}+\frac{1}{c^{2}}\frac{\partial \overrightarrow{E}}{
\partial t}  \, , \nonumber
\end{eqnarray}
where $\rho$ is the density of electric charge of the electric field source, $\vec E$ and $\vec B$ are the electric and magnetic fields respectively, $\epsilon_o$ and $\mu_0$ are the electric permittivity and the magnetic permeability, and $\vec v$ is the velocity of the charges being the source of the magnetic field, and $c$ is the speed of light in vacuum. Equations (\ref{emax}) should be replaced by Bridgman's gravitoelectromagnetic equations, by multiplying the electric field $\vec E$ by $\bar q/m_i$, and the magnetic field $\vec B$ by $\bar q/2m$, where $\bar q /m$ is the charge to mass ratio of a test particle used to detect/measure electric and magnetic fields, cf Fig. (\ref{fig:gemwave}):
\begin{eqnarray}
\nabla \overrightarrow{g_e} &=&\frac{\bar q}{m_i} \frac{1}{ \epsilon_0} \rho \label{emgem} \\
\nabla \overrightarrow{B_{g}} &=&0  \nonumber \\
\nabla \wedge \overrightarrow{g_e} &=&-2\frac{\partial \overrightarrow{B_{g\omega}}}{
\partial t}  \nonumber \\
\nabla \wedge \overrightarrow{B_{g}} &=& \frac{\bar q}{m_i} \frac{\mu_0}{2} \rho
\overrightarrow{v}+\frac{1}{2c^{2}}\frac{\partial \overrightarrow{g_e}}{
\partial t}  \, , \nonumber
\end{eqnarray}
where $\bar q$ and $m_i$ are the electric charge and mass of the test particle respectively, $\vec g_e$ is the electric acceleration of the test particle, and $\vec B_{g}$ is the gravitomagnetic field supported by the test particle, all other symbols are identical to those present in equation (\ref{emax}).
Electromagnetic waves propagating at the speed of light result from equations (\ref{emax}). In a similar manner the propagation of gravitomagnetic waves, propagating at the speed of light, result from equations (\ref{emgem}). However, while the amplitude of the electric field carried by the electromagnetic wave, $E_0$, is independent of the charge to mass ratio $\bar q/m_i$ of the electric-test-mass detecting it:
\begin{equation}
E(x,t)=E_0 \cos\Big(\nu t- \frac{x}{\lambda}\Big)\, ,\label {emwave}
\end{equation}
where $\nu$ and $\lambda$ are the frequency and wavelength of the electromagnetic wave, the amplitude of the gravitomagnetic wave, (which is Bridgman's inertial representation of Maxwell's electromagnetic wave), when interacting with a test particle, does depend on the $\bar q/m_i$ ratio of the test particle:
\begin{equation}
g_e=\frac{\bar q}{m_i} E_0 \cos\Big(\nu t- \frac{x}{\lambda}\Big)\, .\label {qmwave}
\end{equation}
Therefore one given source of electromagnetic waves will generate different (Bridgman) gravitomagnetic waves with varying amplitudes, depending on the different $\bar q/m_i$ ratio of the test charges that are present around the source. This does not forbid the photon to be the quanta of all these different amplitude varying gravitomagnetic waves, since the photon energy only depends on the wave's frequency $\nu$,
\begin{equation}
\epsilon = h \nu\, , \label{Planck1}
\end{equation}
where $h$ is Planck's constant. But does a (Bridgman) gravitomagnetic wave carry energy? An electromagnetic wave does carry energy its flux radiance (W/m$^2$) being proportional to the square of its electric amplitude $E_0$ \cite{Hecht}.
\begin{equation}
I=\frac{c}{2} \frac{1}{4\pi K} E_0^2 \, . \label{Planck2}
\end{equation}
To find the gravitomagnetic analog of the electromagnetic flux radiance we need to substitute, in equation (\ref{Planck2}), the Coulomb constant $K$, by equation(\ref{bri9}), and the electric field $E_0$, by the electric acceleration $\bar q E_0/ m$ (cf equation (\ref{bri12})), to find:
\begin{equation}
I_{g}=\frac{c}{2} \frac{1}{4\pi K} \frac{\bar q}{m_i} E_0^2 \, . \label{Planck3}
\end{equation}
The units of the new quantity $I_{g}$ are (Coulomb . Meter$^2$/ Second$^3$)/(Meter$^2$), which does not correspond to any known physical quantity of the International System of units. Ultimately Bridgman operational description of electromagnetism leads to an unknown physical quantity (Coulomb . Meter$^2$/ Second$^2$) that should replace energy (Joule=Kilogram . Meter$^2$/ Second$^2$) in electromagnetic interactions. The new physical quantity (Coulomb . Meter$^2$/ Second$^2$) is operationally ill defined because there is no experimental procedure to measure it directly without measuring first a force. The illness in the definition of this quantity resides in the fact that the electric charge of a body cannot cause any inertial effect independently of its mass. Since Bridgman's theory of electromagnetism does not lead to a suitable physical quantity to replace energy in the electromagnetic interaction, we conclude that electromagnetic fields cannot depend on the charge to inertial mass ratio of test bodies as prescribed by Bridgman.

\section{Gravitational waves cannot depend on test body's passive-gravitational to inertial mass ratio}
For gravitation it is not electric-acceleration and magnetic-gravitomagnetic fields that propagate, (like it was the case for Bridgman's approach of electromagnetism). Gravitational waves consist in the propagation of spacetime curvature according to Einstein's theory of general relativity. This theory predicts, that the total power emitted \cite{Hobeson}, in the form of gravitational waves, by a spinning dumbbell, of total mass $2M$, and radius $a$, is:
\begin{equation}
P_{GW}=\frac {dE}{dt}=-\frac{32G}{5c^5} I^2 \omega^6 \, , \label{Einstein1}
\end{equation}
where $I=2 Ma^2$ is the moment of inertia of the dumbell, and $\omega$ is its angular velocity. Equation (\ref{Einstein1}) has been verified experimentally to a high accuracy \cite{Hulse}. To apply Bridgman's views of gravitational fields to Einstein's spacetime metric theory of gravitation, we must substitute, in equation (\ref{Einstein1}), Newton's gravitational constant $G$  by equation (\ref{bri4}). Hence equation (\ref{Einstein1}) becomes:
\begin{equation}
P_{GW}=\frac {dE}{dt}=-\frac{m_p}{m_i} \frac{32G}{5c^5} I^2 \omega^6 \, .\label{Einstein2}
\end{equation}
Equation (\ref{Einstein2}) tells us that the total power emitted by a gravitational wave source, in the form of gravitational waves, would depend on the passive gravitational mass-to inertial mass ratio $m_p/m_i$ of the test masses detecting the gravitational wave (far away from the source). Hence equation (\ref{Einstein2}) does not violate energy conservation and the laws of causality only if:
\begin{equation}
\frac{m_p}{m_i}=1 \, . \label{Einstein3}
\end{equation}
Therefore gravitational waves, which have been widely confirmed since their first observation in 2015 \cite{Abbott}, cannot depend on the test body's gravitational to inertial mass ratio, this can only happen when this ratio is exactly equal to 1! If it would be different from 1, different test masses would associate different gravitational energies to the same gravitational wave source, this would be a clear violation of energy conservation. It would also represent a violation of causality because the gravitational energy of the source  would only be determined after it is detected by a test body (since the test body's gravitational to inertial mass ratio would contribute to determine the gravitational energy of the source). Therefore gravitational fields and spacetime curvature cannot depend on the test body's passive-gravitational to inertial mass ratio, as prescribed by Bridgman!

\section*{Conclusions}
We demonstrated that the operational Bridgman's approach to electromagnetic and gravitational interactions, requiring that the fundamental physical nature of electromagnetic and gravitational fields reduce ultimately to the forces experienced by the test bodies, is not sustainable. For the case of gravitation it leads to the violation of the law of conservation of energy, and the law of causality. For what concerns electromagnetism, it requires operationally ill defined quantities (Coulomb . Meter$^2$/ Second$^2$) that would replace the energy exchanged between two bodies when interacting electromagnetically. This results also disqualifies Murat Ozer's theory of unification of electromagnetism and gravitation \cite{MO1} \cite{MO2} \cite{MO3}, which is founded on the assumption that electromagnetic fields do depend on the charge to mass ratio of test bodies, and spacetime curvature depends on the passive gravitational to inertial mass ratio of test bodies, and which assumes that it is possible to have different values (different from 1) for the passive gravitational-to inertial mass ratio $m_p/m_i$ of test bodies (without disrupting Einstein's theory of general relativity).

\end{document}